\documentclass[12pt,english,aps,preprint]{revtex4}
\usepackage[T1]{fontenc}
\usepackage[latin1]{inputenc}
\usepackage{babel}
\usepackage{graphics}

\makeatletter

\providecommand{\LyX}{L\kern-.1667em\lower.25em\hbox{Y}\kern-.125emX\@}

\newcommand\dslash{\makebox[1.5pt][l]{D}\slash\ }
\def\Tr   {\mathop{\hbox{Tr}}}
\def\Det   {\mathop{\hbox{Det}}}
\def\Real   {\mathop{\hbox{Re}}}

\makeatother
\begin{document}
{\raggedleft COLO-HEP-482\par}

{\raggedleft July 2002\par}

{\centering \textbf{\Large Partial-Global Stochastic Metropolis Update
for Dynamical Smeared Link Fermions} {\Large }\Large \par}

\author{Andrei Alexandru}

\email{alexan@pizero.colorado.edu}

\thanks{}

\affiliation{Department of Physics, University of Colorado, Boulder, CO-80304-0390}

\author{Anna Hasenfratz}

\email{anna@eotvos.colorado.edu}

\thanks{}

\affiliation{Department of Physics, University of Colorado, Boulder, CO-80304-0390}

\begin{abstract}
We performed dynamical simulations with HYP smeared staggered fermions
using the recently proposed partial-global stochastic Metropolis algorithm
with fermion matrix reduction and determinant breakup improvements.
In this paper we discuss our choice of the action parameters and study
the autocorrelation time both with four and two fermionic flavors
at different quark mass values on approximately \( 10 \) fm\( ^{4} \)
lattices. We find that the update is especially efficient with two
flavors making simulations on larger volumes feasible. 

\end{abstract}
\maketitle

\section{introduction}

Smeared link actions, actions where the traditional thin link gauge
connection is replaced by some sort of smeared or fattened link, have
gained popularity in recent years. Smearing removes the most ultra-violet
gauge field fluctuations improving chiral and flavor symmetry in Wilson/clover
and staggered fermion actions \cite{Hasenfratz:2001hp, Orginos:1998ue, Orginos:1999cr, Bernard:2001av, Hasenfratz:2002}.
These actions also have better scaling and topological properties.
While smearing is straightforward to implement on quenched configurations,
the dynamical simulation of smeared actions is far from trivial. If
the smearing is linear in the original gauge variables,  numerical
updates based on the molecular dynamics evolution is complicated but
possible. On the other hand, if the smeared links are projected back
to the gauge group and therefore are not linear in the thin link variables,
algorithms that require the calculation of the fermionic force do
not work.

In a recent paper \cite{Hasenfratz:2002} we suggested that smeared
link actions can be effectively simulated with partial-global heatbath
or over-relaxation updates combined with a Metropolis type accept-reject
step that requires only a stochastic estimate of the fermionic determinant.
In a subsequent publication \cite{Hasenfratz:2002ym} we studied the
stochastic evaluation of the fermionic determinant. The most important
conclusion of Ref. \cite{Hasenfratz:2002ym} is that the standard
deviation of the stochastic estimator can be very large, even divergent,
thus introducing unacceptably large autocorrelation times in simulations.
We suggested two ways to improve the stochastic estimator. In this
paper we apply those improvements in two and four flavor dynamical
staggered action simulations and study the effectiveness of the partial-global
stochastic Metropolis (PGSM) algorithm. Since the standard deviation
of the stochastic estimator depends on what fraction of the original
configuration is changed in one partial-global update, we investigate
the autocorrelation time as the function of the number of links touched
in the partial-global heatbath step. This knowledge is necessary to
optimize the parameters of the PGSM algorithm and compare its effectiveness
to other simulation methods. In our numerical simulation we use hypercubic
smeared (HYP) links \cite{Hasenfratz:2001hp} with staggered fermions,
though many of the concepts and improvements of the PGSM algorithm
can be directly applied to other kind of smeared links and fermionic
formulations. 

To make this paper self-contained, in Sect. 2. we briefly summarize
our smeared link action, the PGSM method and its improvements. In
Sect. 3. we discuss the details of the updating and the parameters
of the simulations. Sect. 4. contains our autocorrelation time results.
A short conclusion completes this paper.

\section{The Smeared Link Action and the Partial-global Update }

\subsection{The action}

We consider a smeared link action of the form\begin{equation}
\label{full_action}
S=S_{g}(U)+\bar{S}_{g}(V)+S_{f}(V),
\end{equation}
 where \( S_{g}(U) \) and \( \bar{S}_{g}(V) \) are gauge actions
depending on the thin links \( \{U\} \) and smeared links \( \{V\} \),
respectively, and \( S_{f} \) is the fermionic action depending on
the smeared links only. We assume that the smeared links are constructed
deterministically. In our numerical simulation we use hypercubic (HYP)
blocking. The HYP links are optimized non-perturbatively to be maximally
smooth, but they are also tree-level perturbative improved. The construction
and properties of HYP smearing are discussed in detail in Ref. \cite{Hasenfratz:2001hp}.

We use a plaquette gauge action for \( S_{g}(U) \)\begin{equation}
\label{S_g(U)}
S_{g}(U)=-\frac{\beta }{3}\sum _{p}\Real \Tr (U_{p}),
\end{equation}
 and we will choose \( \bar{S}_{g}(V) \) in Sect. II.D to improve
computational efficiency. The fermionic action describing \( n_{f} \)
degenerate flavors with staggered fermions is \begin{equation}
S_{f}(V)=-\frac{n_{f}}{4}\ln \, \Det \, \Omega (V)
\end{equation}
 with \begin{equation}
\label{Omega}
\Omega (V)=(M^{\dagger }(V)M(V))_{\rm {even,even}}
\end{equation}
defined on even sites only. \( M(V) \) in Eq. \ref{Omega} is the
standard staggered fermion matrix.

\subsection{The partial-global stochastic Metropolis update}

In Refs. \cite{Knechtli:2000ku, Hasenfratz:2002, Hasenfratz:2002ym}
a partial-global stochastic updating (PGSM) algorithm was developed
to simulate smeared link actions. First a subset of the thin links
\( \{U\} \) are updated and a new thin gauge link configuration \( \{U'\} \)
is proposed. The transition probability \( p(U,U') \) of the update
satisfies detailed balance with \( S_{g}(U) \) \begin{equation}
p(U,U')e^{-S_{g}(U)}=p(U',U)e^{-S_{g}(U')}.
\end{equation}
 The proposed configuration is accepted with the probability \begin{equation}
\label{Pstoch}
P_{\rm {acc}}=\rm {min}\{1,e^{-\bar{S}_{g}(V')+\bar{S}_{g}(V)}e^{-\xi ^{*}[\Omega ^{-1}(V')\Omega (V)-1]\xi }\},
\end{equation}
 where the stochastic vector \( \xi  \) is generated with Gaussian
distribution, \( \rm {exp}(-\xi ^{*}\xi ) \). The PGSM algorithm
satisfies detailed balance \cite{Grady:1985fs}. A somewhat similar
updating method has been proposed and investigated in Refs. \cite{Lin:1999qu, Joo:2001bz}.
The Kentucky Noisy Monte Carlo method proposes gauge fields created
with a pure gauge global update but the accept reject step is based
on the noisy evaluation of the fermionic determinant. So far that
method has been tested with Wilson fermions at rather heavy quark
masses.

\subsection{Improving the partial-global stochastic Metropolis update}

The PGSM algorithm averages the stochastic estimator of the fermionic
determinant together with the gauge configuration ensemble and it
can be efficient only if the standard deviation of the stochastic
estimator is small. The standard deviation of the stochastic estimator
as described by Eq. \ref{Pstoch} is governed by the inverse determinant
of the matrix \( (2\Omega ^{-1}(V')\Omega (V)-1) \) and diverges
if even one of the eigenvalues of \( \Omega ^{-1}(V')\Omega (V) \)
is less than 1/2 \cite{Hasenfratz:2002ym}. One can control the standard
deviation by reducing the spread of the eigenvalues of the fermionic
matrix \( \Omega (V) \). In Ref. \cite{Hasenfratz:2002ym} we considered
the combination of two separate methods to achieve that. Here we only
briefly summarize them.

\begin{enumerate}
\item \textbf{Reduction:} The fermionic matrix reduction \cite{Hasenbusch:1998yb, Knechtli:2000ku}
removes the most UV part of the fermionic operator by defining a reduced
matrix \begin{equation}
\Omega _{r}=\Omega e^{-2f(\Omega )}
\end{equation}
 and rewriting the fermionic action as \begin{equation}
\label{S_f_mod}
S_{f}(V)=-\frac{n_{f}}{4}\Big (\ln \, \Det \, \Omega _{r}(V)+2\Tr f(\Omega )\Big ).
\end{equation}
 If the function \( f \) is a polynomial of \( \Omega  \), the trace
of \( f(\Omega ) \) can be evaluated exactly and only the determinant
of the reduced matrix \( \Omega _{r} \) has to be calculated stochastically.
The parameters of \( f \) are chosen such as to minimize the eigenvalue
distribution of the reduced fermionic matrix \( \Omega _{r} \). In
Ref. \cite{Hasenfratz:2002ym} we optimized \( f(\Omega ) \) for
staggered fermions assuming a simple eigenvalue distribution derived
in the free field limit. The optimization procedure is easy to implement
for any other fermionic action but the optimized coefficients will
be different. For nearest neighbor staggered fermions with the choice
\begin{equation}
\label{f_omega}
f(\Omega )=-0.34017+0.35645\Omega -0.030379\Omega ^{2}+0.000957\Omega ^{3}
\end{equation}
 the conditioning number of the fermionic matrix is reduced by about
a factor of 30. The smallest possible eigenvalue of the operator \( \Omega ^{-1}(V')\Omega (V) \)
is now about \( 8(am)^{2} \). This is a significant improvement but
it is not sufficient to guarantee a small or even finite standard
deviation of the stochastic estimator. That can be achieved by an
additional improvement step, the determinant breakup. 
\item \textbf{Determinant breakup}: Writing the fermionic action in the
form \begin{equation}
\label{S_f_mod2}
S_{f}(V)=-\frac{n_{f}}{4}\Big (n_{b}\ln \, \Det \, \Omega ^{1/n_{b}}_{r}(V)+2\Tr f(\Omega )\Big )
\end{equation}
 suggests that the stochastic part of the estimator can be evaluated
using \( n_{b}n_{f}/4 \) independent Gaussian vectors as\begin{equation}
\label{P_acc_mod}
P_{\rm {acc}}=\rm {min}\Big \{1,e^{-\Delta \bar{S}_{g}+\frac{n_{f}}{2}\Delta f}\, e^{-\sum ^{n_{b}n_{f}/4}_{i=1}\xi _{i}^{*}[\Omega _{r}^{-1/2n_{b}}(V')\Omega _{r}^{1/n_{b}}(V)\Omega _{r}^{-1/2n_{b}}(V')-1]\xi _{i}}\Big \}.
\end{equation}
Here \( \Delta \bar{S}_{g}=\bar{S}_{g}(V')-\bar{S}_{g}(V) \), \( \Delta f=\Tr f(\Omega ')-\Tr f(\Omega ) \),
and the argument of the stochastic estimator is written in an explicitly
Hermitian form. The standard deviation of the stochastic estimator
is greatly reduced with the determinant breakup, it is now finite
as long as the matrix \( \Omega _{r}^{-1}(V')\Omega _{r}(V) \) has
no eigenvalue smaller than \( 1/2^{n_{b}} \). With non-zero quark
mass \( n_{b} \) can always be chosen such that this condition is
satisfied. Approximating the lowest eigenvalue of \( \Omega _{r}^{-1}(V')\Omega _{r}(V) \)
as \( 8(am)^{2} \), \( n_{b}=4 \) and 8 is barely sufficient with
\( am=0.1 \) and 0.04. A safer choice is to use \( n_{b}=8 \) and
12, respectively.
\end{enumerate}
The determinant breakup reduces the statistical fluctuations of the
stochastic estimator by taking the sum of \( n_{b}n_{f}/4 \) smaller
terms instead of \( n_{f}/4 \) original terms in the exponent. While
taking the average of several stochastic estimates in the acceptance
step of the original PGSM algorithm (Eq. \ref{Pstoch}) violates the
detailed balance condition, the determinant breakup procedure is still
exact.

An added bonus of the determinant breakup is that now simulating arbitrary
number of degenerate or non-degenerate flavors is straightforward
as long as \( n_{b} \) is chosen to be a multiple of 4.

\subsection{The smeared link action with reduced fermion matrix}

The reduction of the fermionic matrix is compensated by an effective
smeared link gauge action \( -\frac{n_{f}}{2}\Tr f(\Omega ) \). With
the choice of \( f \) a cubic polynomial and the standard nearest
neighbor staggered action \( \Tr f \) is the sum of 4- and 6-link
gauge loops and can be evaluated exactly \cite{Hasenfratz:2002ym}.
However, even this calculation can be avoided if we choose \( \bar{S}_{g}(V) \)
so it cancels \( \Tr f(\Omega ) \). With the choice\begin{equation}
\bar{S}_{g}(V)=\frac{n_{f}}{4}\Big (2\Tr f(\Omega )-\frac{\gamma }{3}\sum _{p}\Real \Tr (V_{p})\Big )
\end{equation}
 the dynamical action simplifies to \begin{equation}
\label{S_final}
S=-\frac{\beta }{3}\sum _{p}\Real \Tr (U_{p})-\frac{\gamma }{3}\frac{n_{f}}{4}\sum _{p}\Real \Tr (V_{p})-\frac{n_{f}}{4}\ln \, \Det \, \Omega _{r}(V)
\end{equation}
 with \( \beta  \) and \( \gamma  \) two tunable gauge action parameters.
In the PGSM algorithm the new gauge configuration is proposed with
the thin link pure gauge action and is accepted with probability\begin{eqnarray}
P_{\rm {acc}} & = & \rm {min}\Big \{1,e^{-\Delta S}\Big \},\label{P_acc_final} \\
\Delta S & = & -\frac{\gamma n_{f}}{12}\sum _{p}(\Real \Tr (V'_{p})-\Real \Tr (V_{p}))\label{Delta_S} \\
 & + & \sum ^{n_{b}n_{f}/4}_{i=1}\xi _{i}^{*}[\Omega _{r}^{-1/2n_{b}}(V')\Omega _{r}^{1/n_{b}}(V)\Omega _{r}^{-1/2n_{b}}(V')-1]\xi _{i}.\nonumber 
\end{eqnarray}

Eqs. \ref{S_final} and \ref{P_acc_final} give the final form of
our action and the stochastic estimator. One should note that the
action defined in Eq. \ref{S_final} depends on the parameters of
the reduction function \( f \). The dependence is weak, and since
the coefficients in \( f \) are fixed, they become negligible in
the continuum limit. 

When expressed in terms of gauge loops, \( \Tr f(\Omega ) \) contains
4- and 6-link loops. On \( N_{T}=4 \) and 6 lattices that includes
Polyakov lines, loops wrapping around the temperature direction of
the lattice. In Refs. \cite{Knechtli:2000ku, Hasenfratz:2002, Hasenfratz:2002ym}
we suggested to remove these terms by explicitly including them in
\( \bar{S}_{g}(V) \). This removal is by no means necessary but might
influence finite volume effects.

\section{Numerical simulations with the HYP action}

\subsection{Tuning the simulation parameters}

Our final gauge action has three free parameters: in addition to the
quark mass, there are two gauge couplings, \( \beta  \) and \( \gamma  \),
that can be tuned independently. Since the term of the action proportional
to \( \gamma  \) is included in the acceptance step, we want to keep
it small. Also, a finite \( \gamma  \) coefficient in the continuum
\( \beta \to \infty  \) limit can be neglected and perturbative results
that were obtained with \( \bar{S}_{g}(V)=0 \) remain valid. On the
other hand a negative \( \gamma  \) coupling allows the increase
of the coupling \( \beta  \) in \( S_{g}(U) \), compensating for
the renormalization of the gauge coupling due to the fermions. Ideally
one would like to choose \( \beta  \) such that the gauge action
proposed with \( S_{g}(U) \) matches the configurations of the dynamical
action either at long distance (lattice spacing), short distance (plaquette),
or, ideally both. We found that choosing the coefficient of the thin
link pure gauge action somewhat smaller than what would match the
desired lattice spacing of the dynamical system matches the plaquette
expectation values and provides a good parameter set for the dynamical
system.
\begin{table}
\begin{tabular}{|c|c|c|c|c|}
\hline 
\( \beta  \)&
\( \gamma  \)&
\( am \)&
\( a \){[}fm{]}&
\( <\Tr U_{p}> \) \\
\hline
\hline 
5.65&
-0.10&
0.10&
0.165(1)&
1.6212(4)\\
\hline 
5.65&
-0.14&
0.06&
0.161(1)&
1.6216(5)\\
\hline 
5.65&
-0.15&
0.04&
0.160(1)&
1.6216(4)\\
\hline 
5.65&
0&
\( \infty  \) &
\( \sim  \)0.19 &
1.6128(3)\\
\hline
\end{tabular}

\caption{The parameters and lattice spacing of the \protect\( n_{f}=2\protect \)
runs. The last column shows the matching of the plaquette expectation
values between the dynamical and pure gauge (\protect\( am=\infty \protect \))
simulations. \label{Table_params}}
\end{table}

\begin{table}
\begin{tabular}{|c|c|c|c|c|}
\hline 
\( \beta  \)&
\( \gamma  \)&
\( am \)&
\( a \){[}fm{]}&
\( <\Tr U_{p}> \) \\
\hline
\hline 
5.65&
-0.10&
0.10&
0.173(2)&
1.6033(4)\\
\hline 
5.65&
-0.15&
0.04&
0.165(3)&
1.5913(4)\\
\hline 
5.65&
0&
\( \infty  \) &
\( \sim  \)0.19 &
1.6128(3)\\
\hline
\end{tabular}

\caption{Same as table \ref{Table_params} but for the \protect\( n_{f}=4\protect \)
flavor runs.\label{Table_params_4}}
\end{table}
 Table \ref{Table_params}, where we collected the parameter values
of our \( n_{f}=2 \) flavor runs, illustrates this point. All simulations
were done on \( 8^{3}\times 24 \) lattices and we tried to tune the
lattice spacing, measured by the Sommer parameter \( r_{0} \) \cite{Sommer:1994ce},
to be \( a=0.16-0.17 \) fm. We chose the coupling of \( S_{g}(U) \)
to be \( \beta =5.65 \) and kept it fixed, while tuned the coefficient
\( \gamma  \) with the quark mass. As the last column of table \ref{Table_params}
shows, the plaquette expectation value is almost the same on the dynamical
and pure gauge configurations. The matching works similarly in case
of \( n_{f}=4 \) flavors as table \ref{Table_params_4} illustrates.
The pion to rho mass ratios for both the two and four flavor runs
vary between 0.55 and 0.70, consistent with a renormalization factor
\( Z_{m}\approx 1 \).

The matching of the plaquette expectation value with the parameters
of tables \ref{Table_params},\ref{Table_params_4} are almost perfect,
but the lattice spacing between the dynamical and pure gauge runs
differ by 10-15\%. Our attempt to increase the thin link gauge coupling
to \( \beta =5.7 \) and decrease \( \gamma  \) to achieve the same
lattice spacing failed. As we decreased \( \gamma  \) the simulations
showed sudden, large fluctuations, signaling perhaps a phase transition.
We have not yet explored the full phase diagram, rather we decided
to use a somewhat mismatched coupling.

Since the term \( 2\Tr f(\Omega ) \) in \( \bar{S}_{g}(V) \) compensates
for some of the gauge coupling renormalization, the coefficient \( \gamma  \)
is fairly small at every quark mass we considered. At different \( \beta  \)
couplings and mass values one could choose different \( \gamma  \)
couplings. However as the role of the \( \gamma  \) term in the action
is to compensate for the renormalization of the pure gauge coupling
due to the dynamical quark mass, it is natural to fix \( \gamma  \)
as the function of the quark mass, independent of \( \beta  \). As
the shift of the gauge coupling between pure gauge and dynamical actions
remains finite  at \( am=0 \), one can choose \( \gamma  \) finite
and small even in the chiral limit.

\subsection{The polynomial approximation}

To assure the finiteness of the standard deviation of the stochastic
estimator it is imperative to break up the determinant as in Eq. \ref{S_f_mod2}.
With quark masses \( am=0.1-0.04 \) we need \( n_{b}=8-12 \) determinant
break up to keep the standard deviation small. That implies that in
the stochastic estimator we need the \( n_{b}^{th} \) root of the
fermionic matrix, the \( 2n_{b}^{th} \) root of its inverse. To calculate
the \( n^{th} \) root of the matrices involved we use a polynomial
approximation. This approach was originally developed to approximate
the inverse of the fermion matrix \cite{Montvay:1996ea, Montvay:1997vh}
and has been used extensively with the polynomial Hybrid Monte Carlo
method. In our case it is most efficient to approximate the appropriate
power of the reduced fermionic matrix \( \Omega _{r} \) \begin{eqnarray}
P_{k}^{(n_{b})}(\Omega )\simeq  & \Omega _{r}^{-1/2n_{b}} & =\Omega ^{-1/2n_{b}}\exp (f(\Omega )/n_{b}),\nonumber \\
Q^{(n_{b})}_{l}(\Omega )\simeq  & \Omega _{r}^{1/n_{b}} & =\Omega ^{1/n_{b}}\exp (-2f(\Omega )/n_{b}),\label{poly_form} 
\end{eqnarray}
where \( P^{(n_{b})}_{k} \) and \( Q^{(n_{b})}_{l} \) are \( k \)
and \( l \) order polynomials of the fermionic matrix \( \Omega  \).
To accomplish this we determine polynomials \( P^{(n_{b})}_{k} \)
and \( Q^{(n_{b})}_{l} \) that approximate\begin{eqnarray}
P_{k}^{\left( n_{b}\right) }\left( x\right)  & \simeq  & \left( x\, e^{-2f\left( x\right) }\right) ^{-1/2n_{b}},\nonumber \\
Q_{l}^{\left( n_{b}\right) }\left( x\right)  & \simeq  & \left( x\, e^{-2f\left( x\right) }\right) ^{1/n_{b}},
\end{eqnarray}
on the entire spectrum of \( \Omega  \). In general, to determine
a polynomial \( T_{k} \) that approximates a function \( g\left( x\right)  \)
on the interval \( \left[ \lambda _{0},\lambda _{1}\right]  \) we
minimize the ``distance{}'':\begin{equation}
\label{distance}
\delta ^{2}\left( t_{k}\right) =\int _{\lambda _{0}}^{\lambda _{1}}dx\, \rho \left( x\right) \, \left( T_{k}\left( x\right) -g\left( x\right) \right) ^{2},
\end{equation}
with respect to \( t_{k} \), the coefficients of the polynomial \( T_{k} \).
The weight function \( \rho \left( x\right)  \) is chosen in accordance
to the problem's requirements. In our case it will be, ideally, the
spectral density of the operator \( \Omega  \). The advantage of
this procedure is that the minimization turns into a linear problem. 

The first step is to determine the spectral bounds of \( \Omega  \).
The lower bound is \( \lambda _{0}=4m^{2} \), but the upper bound
determination is slightly more complicated. The only firm bound for
\( \dslash ^{2} \) that we are aware of is \( 64 \). However, in
our simulations we have seen that the highest eigenvalue of \( \Omega  \)
on thermalized configurations is around \( \sim 20 \) when \( \Omega  \)
is defined in terms of thin links, and \( \sim 16 \) when \( \Omega  \)
is defined in terms of the HYP links. We have chosen for our upper
bound \( \lambda _{1}=16.4 \) to be on the safe side. As a rule of
the thumb, we observed that the highest eigenvalue for a particular
operator \( \dslash ^{2} \), when defined on the HYP links, is just
slightly higher than the one the operator assumes in the free field
case. Also, it is worth mentioning that if we choose too low a boundary
for our polynomial approximation (lower than the highest eigenvalue
of \( \Omega  \)) the algorithm starts behaving erratically.

Once the bounds are set we focus on the weight function. The best
choice for the weight function is the spectral density of \( \Omega  \).
However, this density is difficult to determine in the general case.
A more reasonable option is to use the spectral density in the free
field case. Using different weight functions we observed that they
do not alter the polynomial coefficients dramatically and, thus, it
is better to choose a weight function that is more convenient rather
than accurate. The functions that we used have the form \( \rho \left( x\right) =x^{\omega } \),
where \( \omega  \) was chosen to cancel the divergent behavior around
\( \lambda _{0} \). This weight function has the advantage that we
can compute some of the integrals resulting from minimizing the distance
function in Eq. \ref{distance} analytically. This is important since
these integrals have to be computed very precisely for the method
to work (the integrals have to be computed with hundreds of digits
of precision).
\begin{figure}
{\centering \resizebox*{12cm}{!}{\rotatebox{-90}{\includegraphics{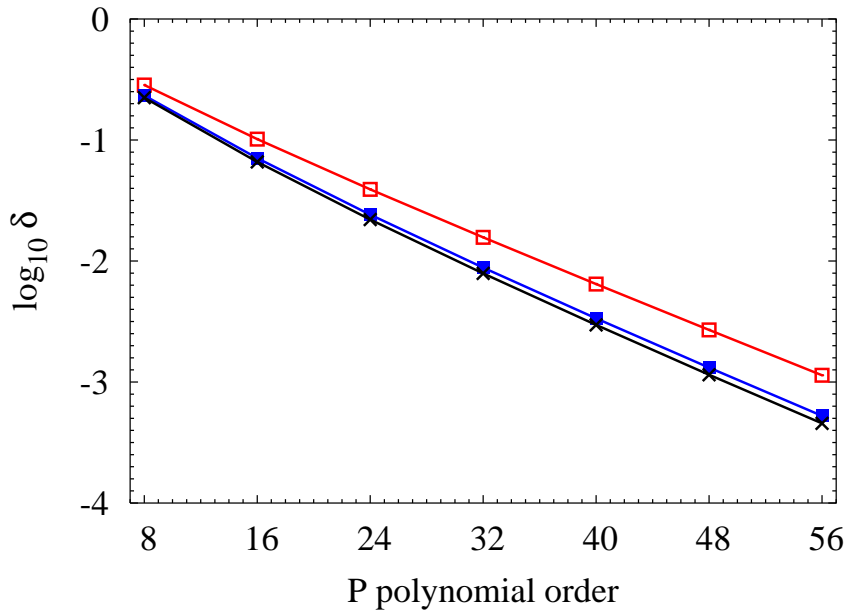}}} \par}

\caption{The accuracy level for the P polynomials with different polynomial
orders and breakup levels: open squares \protect\( n_{b}=1\protect \),
filled squares \protect\( n_{b}=4,\protect \) crosses \protect\( n_{b}=8\protect \)
(\protect\( am=0.10\protect \)).\label{accuracy_P}}
\end{figure}

\begin{figure}
{\centering \resizebox*{12cm}{!}{\rotatebox{-90}{\includegraphics{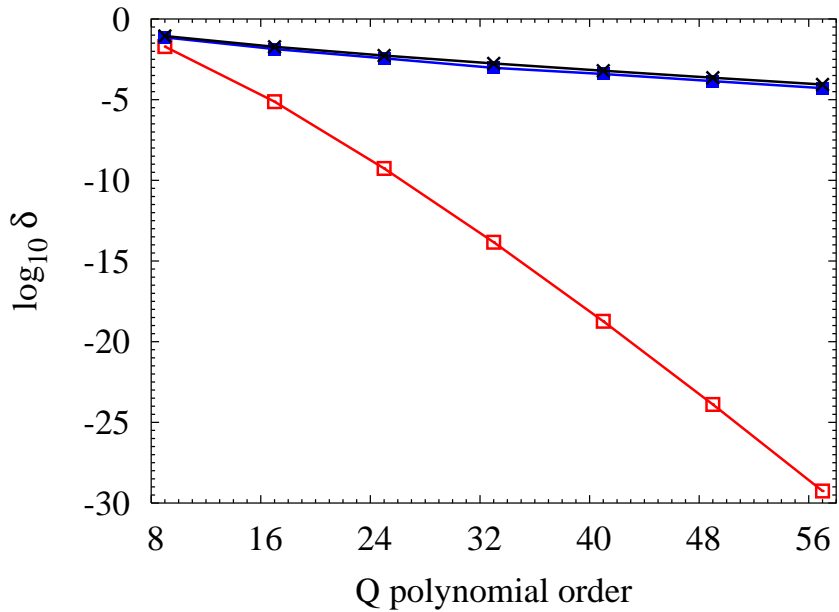}}} \par}

\caption{Same as figure \ref{accuracy_P} but for the Q polynomials. \label{accuracy_Q}}
\end{figure}

In order to get a measure of the accuracy of the polynomial approximation
we used the distance function \( \delta =\sqrt{\delta ^{2}} \) defined
in Eq. \ref{distance} with the weight function set to \( \rho \left( x\right) =1 \).
We have found that it is sufficient to get polynomials with \( \delta \sim 10^{-8} \);
beyond that the roundoff errors become dominant. In figures \ref{accuracy_P}
and \ref{accuracy_Q} we plot the accuracy of the \( \left[ P^{\left( n_{b}\right) }_{k}\right] ^{n_{b}} \)
and \( \left[ Q^{\left( n_{b}\right) }_{l}\right] ^{n_{b}} \) polynomials
for \( n_{b}=1 \), 4 and 8 as the function of the polynomial order
for \( am=0.1 \). The \( n_{b}=1 \) case is of no practical interest,
we show it only to provide a reference. The first thing we notice
is that the level of the approximation for the final functions does
not increase dramatically as we increase the breakup level. This is
startling at first since the resulting polynomial has an order of
\( n_{b}\times k \), with \( k \) the polynomial order. However,
if we think that the level of the approximation is determined by the
number of the coefficients varied in the minimization procedure this
fact is no longer that mysterious. Secondly, we would like to point
out that for \( n_{b}>1 \) the accuracy of the \( P_{k}^{\left( n_{b}\right) } \)
and \( Q_{l}^{\left( n_{b}\right) } \) polynomials with \( k\simeq l \)
is approximately the same. Thus, there is really no point in using
polynomials with \( k \) and \( l \) vastly different since the
level of the accuracy of the final result is going to be dominated
by the smaller order polynomial. From figures \ref{accuracy_P}, \ref{accuracy_Q},
it is clear that in order to achieve the needed approximation level
we have to use polynomials with \( k,l\sim 100-200 \). The necessary
order increases with decreasing quark mass. 

Due to the errors introduced by the polynomials the stochastic estimator
will not be one even when we do not change the configuration at all
(\( \Omega =\Omega ' \)) . This is due to the fact that \( P\left( x\right) Q\left( x\right) P\left( x\right)  \)
is not exactly one. This problem can be largely corrected by rewriting
the estimator as \begin{equation}
\label{stoch_poly}
\xi ^{*}[\Omega _{r}^{-1/2n_{b}}(V')\Omega _{r}^{1/n_{b}}(V)\Omega _{r}^{-1/2n_{b}}(V')-1]\xi \simeq \xi ^{*}P_{k}^{(n_{b})}(\Omega ')[Q_{l}^{(n_{b})}(\Omega )-Q_{l}^{(n_{b})}(\Omega ')]P_{k}^{(n_{b})}(\Omega ')\xi .
\end{equation}
 The right hand side of the expression above has only second order
errors. It is a much better approximation of the quantity on the left
hand side than the estimator that we get by simply replacing \( \Omega _{r} \)
with the polynomial approximation. This step is very important since
the polynomial approximation would otherwise require much larger order
polynomials.

The necessary order for the polynomials \( P \) and Q vary with the
quark mass but we found that in most cases fairly low orders are sufficient.
In simulations with parameters listed in tables \ref{Table_params}
and \ref{Table_params_4} the systematical errors from 128-196 order
polynomials were at the same order as numerical round-off errors.
However this precision is usually unnecessary. All we need to know
for the accept-reject step is whether the stochastic determinant estimate
is smaller or greater than a random number; the precise value is not
important. 

We used a two step approach: we first calculate the stochastic determinant
using low order polynomials and we fall back, i.e recalculate the
stochastic estimator with higher precision, only if the random number
differs from the low order value by less than the estimated error.
This method reduces cost if the fall-back rate is small, less or around
10\%. To determine the optimum small polynomials we vary their order
to minimize the average computational cost. 

We estimate the error the low order polynomials might cause from preliminary
short runs. A typical short run has a couple of hundreds estimator
measurements using the small and the exact (large) polynomials. From
this we determine the error of the small polynomials such that the
values predicted with small and large polynomials agree within this
error at least 99.5\% of the time in the production runs.
\begin{figure}
{\centering \resizebox*{10cm}{!}{\rotatebox{-90}{\includegraphics{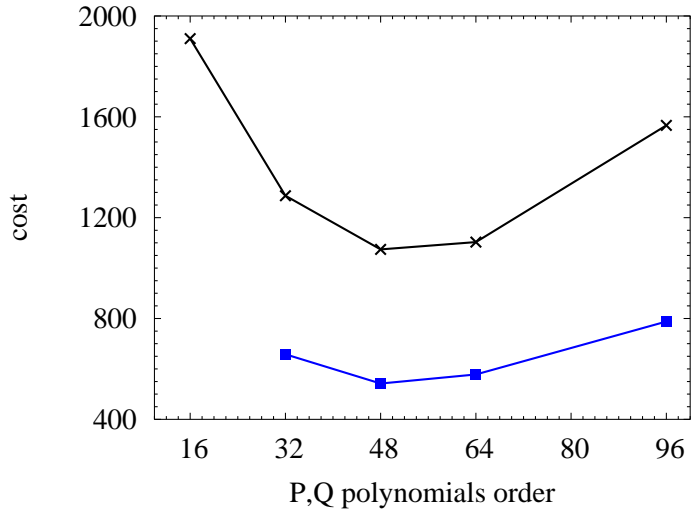}}} \par}

\caption{Average computational cost (number of \protect\( M^{\dagger }M\protect \)
multiplies) for different small polynomials: filled squares \protect\( n_{b}=4\protect \),
crosses \protect\( n_{b}=8\protect \) (\protect\( am=0.04\protect \),
\protect\( k,l\sim 128\protect \) for large polynomials).\label{cost}}
\end{figure}

We found that the order of the optimum small polynomials is independent
of the breakup level, as we can see from figure \ref{cost}, but the
necessary order increases as we decrease the mass. The optimum low
order polynomials we determined to be \( k=24, \) \( l=25 \) for
quark mass \( am=0.1 \), \( k=32 \), \( l=33 \) for \( am=0.06 \)
and \( k=48 \), \( l=49 \) for \( am=0.04 \). The fall-back rate,
i.e. the rate when we had to repeat the stochastic estimator calculation
with higher order polynomials varied between 5\% and 10\%. The outcome
of the accept/reject step changed even less frequently, about 10\%
of the time the fall-back occurred. 

The polynomial approximation is a major part of our simulation. Any
improvement that reduces the necessary polynomial order or the fall-back
rate will immediately reduce the computer time requirements. At present
we are investigating several possibilities along this direction.

\section{The effectiveness of the partial-global stochastic update}

The effectiveness of the PGSM algorithm depends on both the partial-global
heat bath update and the stochastic estimator. The autocorrelation
time of the simulation is the larger of the autocorrelation times
of the heat bath and stochastic steps. If only a small part of the
original links are updated, the fermionic matrix ratio \( \Omega ^{-1}(V')\Omega (V) \)
is close to unity, the standard deviation and the autocorrelation
time of the stochastic estimator is small. On the other hand the heat
bath update is ineffective and has to be repeated many times to update
the configuration. Inversely, if many links are changed at once, the
heat bath update is effective but the standard deviation of the stochastic
estimator increases, increasing the autocorrelation time again. We
would like to find the optimal update where the simulation is most
effective. 

We have measured the integrated autocorrelation time of the plaquette
as the function of the number of links updated at one heat bath step,
\( t_{HB} \), both for \( n_{f}=4 \) and \( n_{f}=2 \) flavor simulations.
The estimates for \( \tau _{\rm {auto}} \) in the following are based
on runs that were 80-100 times longer than \( \tau _{\rm {auto}} \)
itself (except at the largest \( \tau _{\rm {auto}} \) values where
the statistics is smaller). While that does not provide a very good
estimate for the autocorrelation, it is sufficient to distinguish
the heatbath versus stochastic estimator dominated regions.

\begin{figure}
{\centering \resizebox*{12cm}{!}{\rotatebox{-90}{\includegraphics{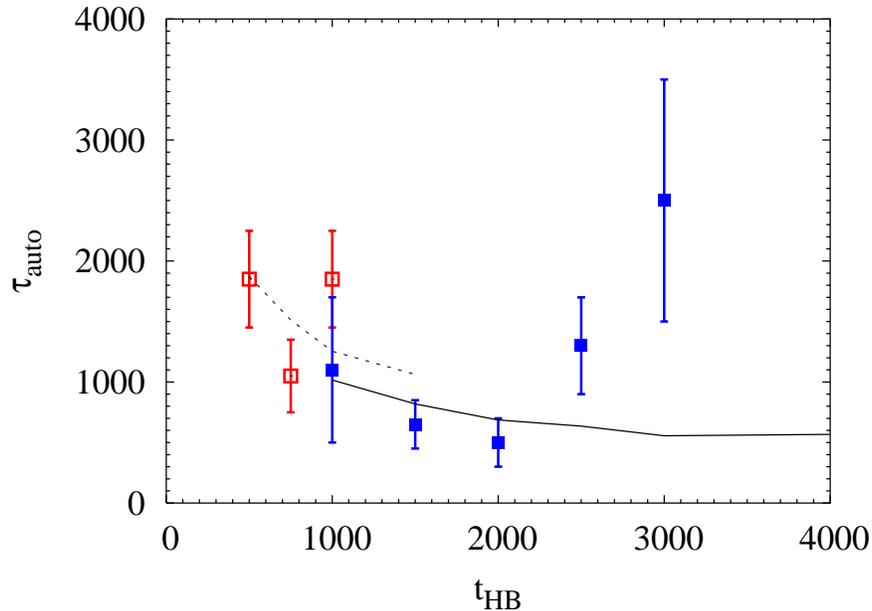}}} \par}

\caption{Autocorrelation of the \protect\( n_{f}=4\protect \), \protect\( am=0.1\protect \)
runs. Open squares: \protect\( n_{b}=4\protect \) , filled squares:
\protect\( n_{b}=8\protect \) determinant breakup. The lines correspond
to the expected autocorrelation based on the heatbath update alone.
\label{tau_nf4}}
\end{figure}
The autocorrelation time of the pure gauge partial-global heat bath
update is easy to estimate. If at each step we update \( t_{HB} \)
links of a lattice of volume V with acceptance rate \( r_{a} \),
the autocorrelation time \( \tau _{pg} \) is related to the autocorrelation
time of the whole lattice pure gauge heat bath update \( \tau _{HB} \)
as \begin{equation}
\label{tau_pg}
\tau _{pg}=\tau _{HB}\frac{4V}{t_{HB}r_{a}}.
\end{equation}
 In our simulations at \( \beta =5.65 \) on \( 8^{3}\times 24 \)
volumes \( \tau _{HB}\sim 10 \). The acceptance rate of the PGSM
depends not only on \( t_{HB} \) but on the determinant breakup parameter
\( n_{b} \) as well. 

The simulation parameters are those listed in tables \ref{Table_params},
\ref{Table_params_4}. We considered \( n_{b}=4 \) and 8 determinant
breakup with the \( n_{f}=4 \), \( am=0.1 \) runs. The open squares
in figure \ref{tau_nf4} show the autocorrelation time for \( n_{b}=4 \),
the filled squares for \( n_{b}=8 \).
\begin{figure}
{\centering \resizebox*{10cm}{!}{\rotatebox{-90}{\includegraphics{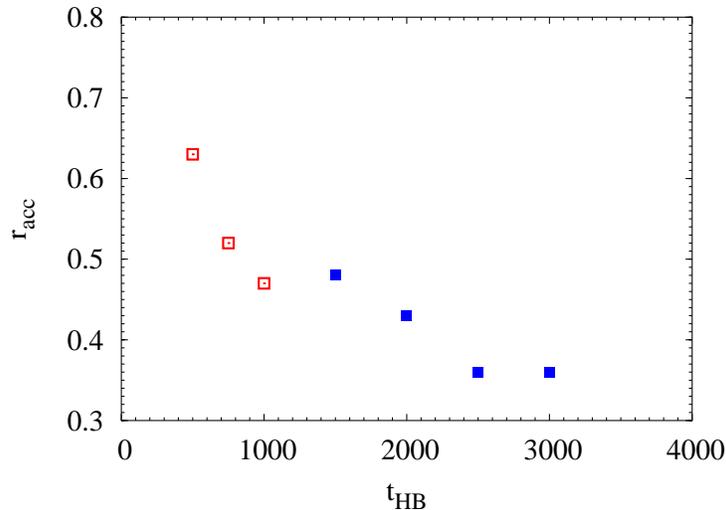}}} \par}

\caption{Acceptance rate of the \protect\( n_{f}=4\protect \), \protect\( am=0.1\protect \)
runs. Notation as in figure \ref{tau_nf4}. \label{accr_nf4}}
\end{figure}
 The dashed line is \( \tau _{pg} \) of Eq. \ref{tau_pg} for \( n_{b}=4 \),
the solid line for \( n_{b}=8 \). For small \( t_{HB} \) values
both \( n_{b}=4 \) and 8 follow the \( \tau _{pg} \) curve, indicating
that the autocorrelation time is dominated by the heatbath step. The
\( n_{b}=4 \) data break off around \( t_{HB}=1000 \), the \( n_{b}=8 \)
around \( t_{HB}=2500 \) signaling that the autocorrelation time
beyond that is dominated by the stochastic estimator. The acceptance
rate, shown in figure \ref{accr_nf4}, varies between 35\% and 65\%
but does not indicate whether the autocorrelation is heatbath or stochastic
estimator dominated. For an effective simulation one should choose
\( t_{HB}=750 \) with \( n_{b}=4 \) or \( t_{HB}=2000 \) with \( n_{b}=8 \).
Which of these is more effective depends on the implementation of
the polynomial approximation in the determinant breakup. In our simulations
updating 2000 links with \( n_{b}=8 \) is the better choice. One
should note that at this mass value the \( n_{b}=4 \) determinant
breakup is just barely adequate to guarantee the finiteness of the
standard deviation of the stochastic estimator, one more reason to
choose \( n_{b}=8 \) in the simulations. 
\begin{figure}
{\centering \resizebox*{12cm}{!}{\rotatebox{-90}{\includegraphics{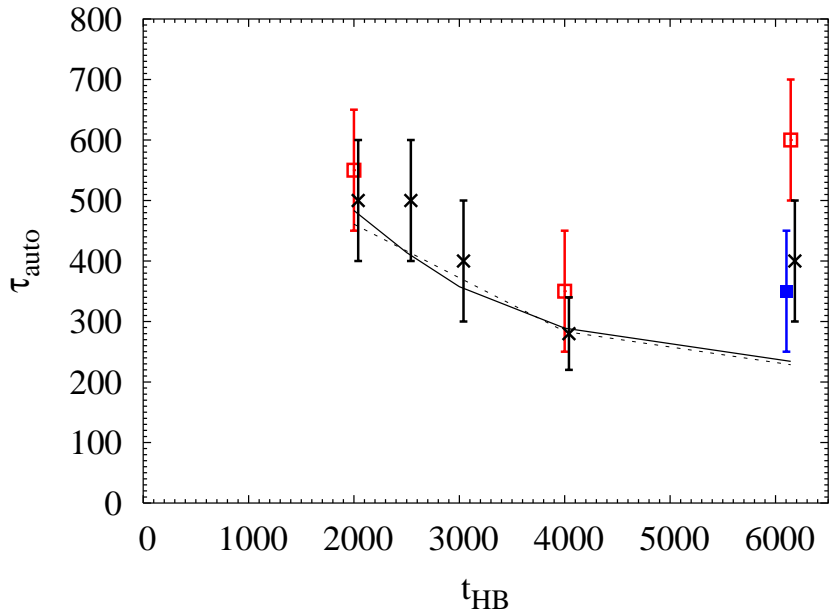}}} \par}

\caption{Autocorrelation of the \protect\( n_{f}=2\protect \) runs. Open
squares: \protect\( n_{b}=8\protect \), \protect\( am=0.1\protect \),
filled squares: \protect\( n_{b}=12\protect \), \protect\( am=0.1,\protect \)
crosses: \protect\( n_{b}=12\protect \), \protect\( am=0.04\protect \)
runs. For clarity the data points are slightly displaced horizontally.
The lines corresponds to the expected autocorrelation based on the
heatbath update alone. Solid line: \protect\( n_{b}=8\protect \),
\protect\( am=0.1\protect \), dashed line: \protect\( n_{b}=12\protect \),
\protect\( am=0.04\protect \). \label{tau_nf2}}
\end{figure}

\begin{figure}
{\centering \resizebox*{10cm}{!}{\rotatebox{-90}{\includegraphics{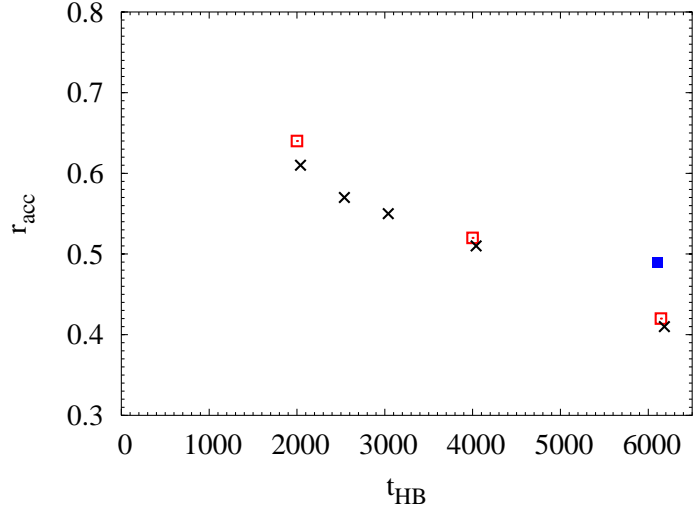}}} \par}

\caption{The acceptance rate of the \protect\( n_{f}=2\protect \) runs. Notation
as in figure \ref{tau_nf2}. \label{accr_nf2}}
\end{figure}

The \( n_{f}=2 \) flavor simulations are considerably more effective.
In figure \ref{tau_nf2}  we plot \( \tau _{\rm {auto}} \) for the
\( am=0.1 \) mass with \( n_{b}=8 \) and \( n_{b}=12 \) and for
the \( am=0.04 \) quark mass with \( n_{b}=12 \) determinant breakup.
Note that the data points with \( t_{HB}=6144 \) touch the maximum
number of links allowed on these \( 8^{3}\times 24 \), close to 10
fm\( ^{4} \) lattices. With both quark mass values only this last
point shows any deviation from the \( \tau _{pg} \) heatbath curve.
The acceptance rate is shown in figure \ref{accr_nf2}. It is interesting
to note that the \( am=0.1 \) runs with \( n_{b}=8 \) determinant
breakup have very similar acceptance rates to the \( am=0.04 \),
\( n_{b}=12 \) runs indicating how the determinant breakup should
increase with decreasing quark mass. We found similar correspondence
in other quantities, like the standard deviation of the stochastic
determinant estimate or the average of the action difference \( \Delta S \).

\begin{figure}
{\centering \resizebox*{12cm}{!}{\rotatebox{-90}{\includegraphics{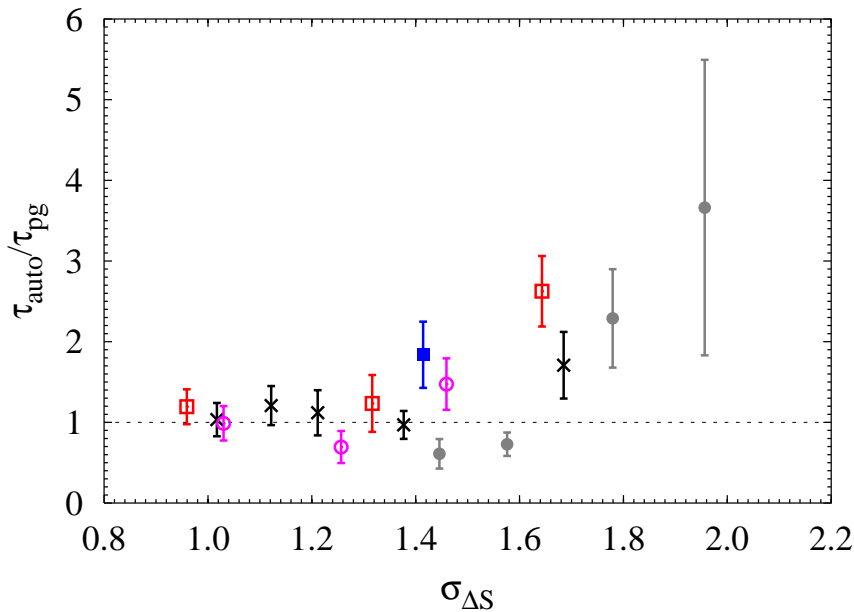}}} \par}

\caption{The autocorrelation time in units of \protect\( \tau _{pg}\protect \)
of Eq. \ref{tau_pg} as the function of the standard deviation of
the stochastic action, \protect\( \Delta S\protect \). Open and filled
circles: \protect\( n_{f}=4\protect \), \protect\( am=0.1\protect \),
\protect\( n_{b}=4\protect \) and 8 runs. Open and filled squares:
\protect\( n_{f}=2\protect \), \protect\( am=0.1\protect \), \protect\( n_{b}=8\protect \)
and 12 runs. Crosses: \protect\( n_{f}=2\protect \), \protect\( am=0.04\protect \),
\protect\( n_{b}=12\protect \) runs. \label{tauotpg}}
\end{figure}

Measuring the autocorrelation time requires long simulations. One
of our goals in this paper is to find an alternate quantity that is
easy to measure but still indicates the crossover from the heatbath
to stochastic estimator regions. We have considered the standard deviation
of the stochastic estimator and also the standard deviation of \( \Delta S \)
as defined in Eq. \ref{Delta_S}. We decided to use the latter as
\( \Delta S \) has an almost Gaussian distribution and its standard
deviation is easier to estimate. In figure \ref{tauotpg} we show
both the \( n_{f}=4 \) and \( n_{f}=2 \) data for the autocorrelation,
measured in units of \( \tau _{pg} \), as the function of \( \sigma _{\Delta S}=\sqrt{<\Delta S^{2}>-<\Delta S>^{2}} \).
As we have already expected from figures \ref{tau_nf4} and \ref{tau_nf2},
most data points are consistent with \( \tau _{\rm {auto}}/\tau _{pg}=1 \),
i.e. heatbath dominated. We see a gradual increase only at \( \sigma _{\Delta S}\geq 1.6 \).
Based on figures \ref{accr_nf4},\ref{accr_nf2} and \ref{tauotpg}
we conclude that the PGSM update is heatbath dominated if the acceptance
rate is around 50\% or higher and the standard deviation of the action
difference \( \sigma _{\Delta S} \) is less than about 1.6. 

\begin{figure}
{\centering \resizebox*{12cm}{!}{\rotatebox{-90}{\includegraphics{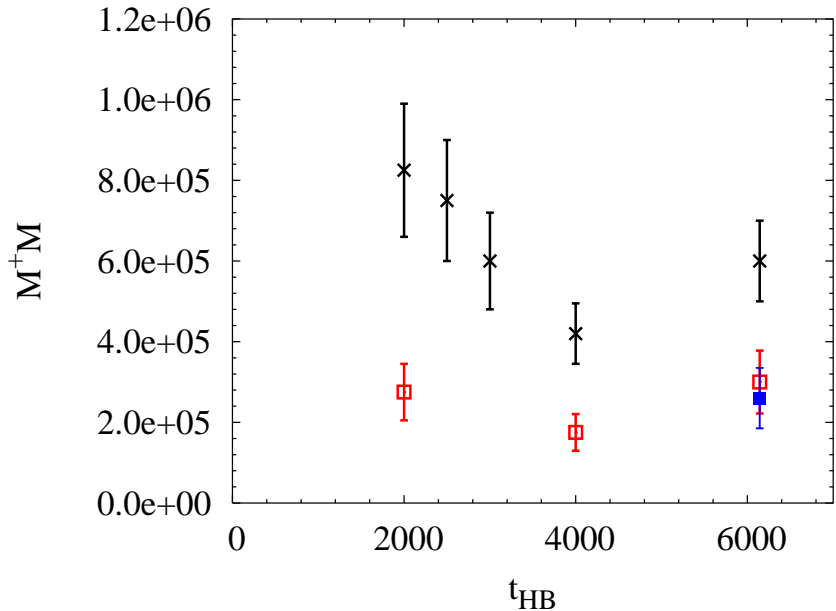}}} \par}

\caption{Cost of creating independent configurations, measured in fermionic
matrix multiplies. Notation is the same as in figure \ref{tau_nf2},
open squares: \protect\( n_{b}=8\protect \), \protect\( am=0.1\protect \),
filled square: \protect\( n_{b}=12\protect \), \protect\( am=0.1,\protect \)
crosses: \protect\( n_{b}=12\protect \), \protect\( am=0.04\protect \)
runs.. \label{M+M_cost}}
\end{figure}

We conclude this section by translating the autocorrelation time measurements
to computer time requirements. In figure \ref{M+M_cost} we plot the
computer time measured in fermionic matrix \( M^{\dagger }M \) multiplies
to create configurations in the \( n_{f}=2 \) flavor simulations
that are separated by \( 2\tau _{\rm {auto}} \) update steps. These
estimates strongly depend on the polynomial approximation but still
give a fair indication of the expected computer cost. One should note
that in figure  \ref{M+M_cost} we consider the time requirement of
the stochastic estimator only. Updating the gauge field and evaluating
the HYP smeared links can result in a 10-20\% overhead on smaller
lattices. In Ref. \cite{Hasenfratz:2002} it was found that a HMC
thin link staggered fermion update with \( n_{f}=4 \) flavors at
similar physical parameter values to our \( am=0.1 \) run requires
about \( 1.2\times 10^{6} \) \( M^{\dagger }M \) multiplies to create
independent configurations . A two flavor run is faster but even than
it is evident that on these 10 fm\( ^{4} \) lattices it is actually
faster to create HYP smeared configurations with PGSM than think link
ones with a small step-size algorithm. 

The autocorrelation measurements and computer time estimates presented
above were based on runs on 10 fm\( ^{4} \) lattices. The PGSM algorithm
scales with the square of the lattice volume. To repeat the above
measurements on 100 fm\( ^{4} \) lattices would increase the computer
time by 100. A factor of 10 increase is due to the increased volume,
and the other factor of 10 is coming from the increased autocorrelation
time. Simulation cost at smaller lattice spacing but same physical
volume increases only linearly with the volume as the number of links
that can be effectively touched increases accordingly.

\section{Conclusion}

In this paper we discussed the practical implementation of the partial-global
stochastic Metropolis (PGSM) algorithm for HYP link staggered fermions.
The PGSM is a two step algorithm where a partial set of the original
gauge links are updated with a heatbath or overrelaxed step and this
proposed configuration is accepted or rejected according to a stochastic
estimate of the fermionic determinant. Even though only a stochastic
estimator is used in the Metropolis step, the algorithm satisfies
detailed balance. 

The effectiveness of the algorithm depends on the matching of the
dynamical action and the pure gauge action used in the first step
of the algorithm, and on the stochastic estimator used in the Metropolis
accept-reject step. The first condition can be met by separating the
pure gauge action into a part that depends on the thin gauge links
and matches the short and/or long distance properties of the dynamical
action, and a part that depends on the smeared links and compensates
for the renormalization of the pure gauge coupling due to the dynamical
fermions. Only the former is used to propose the new configuration,
the latter is included in the accept-reject step. The stochastic estimator
can be significantly improved by fermionic matrix reduction and determinant
breakup. Our \( n_{f}=2 \) flavor simulations indicate that with
these improvements up to 1 fm\( ^{4} \) section of the lattice can
be updated at one time. The update has to be repeated enough times
to refresh the whole lattice, but beyond that the properties of the
heatbath update determine the autocorrelation time. Combining the
heatbath update with overrelaxation could provide an even more effective
algorithm that decorrelates the configurations much faster than the
traditional small step size algorithms. 

\begin{acknowledgments}
We are indebted to F. Knechtli and U. Wolf for clarifying the proof
of the detailed balance condition for the PGSM algorithm. This computation
was carried out on the 32-node Beowulf cluster of the University of
Colorado HEP Theory Group. Our computer code is based on the publicly
available MILC collaboration software.
\end{acknowledgments}
\bibliographystyle{apsrev}
\bibliography{paper}

\end{document}